\documentclass[showpacs]{revtex4}

\usepackage{graphicx}
\usepackage{wrapft}
\usepackage{amsmath,amssymb}
\usepackage{bm}
\usepackage{color}
\usepackage{textcomp}
\usepackage{multirow}
\usepackage{braket}

\begin{document}

\title{Be-$\alpha$ correlations in the linear-chain structure of C isotopes}

\author{Tadahiro Suhara and Yoshiko Kanada-En'yo}

\affiliation{Department of Physics, Graduate School of Science, Kyoto University, Kyoto 606-8502, Japan}

\begin{abstract}
We investigate the linear-chain structures in highly excited states of $^{14}$C using a generalized molecular orbital model, 
by which we incorporate an asymmetric configuration of three $\alpha$ clusters in the linear-chain states. 
By applying this model to the $^{14}$C system, we study the $^{10}$Be+$\alpha$ correlation in the linear-chain state of $^{14}$C. 
To clarify the origin of the $^{10}$Be+$\alpha$ correlation in the $^{14}$C linear-chain state, 
we analyze linear $3\alpha$ and $3\alpha$+$n$ systems in a similar way.
We find that a linear 3$\alpha$ system prefers the asymmetric 2$\alpha$+$\alpha$ configuration,
whose origin is the many body correlation incorporated by the parity projection.
This configuration causes an asymmetric mean field for two valence neutrons,
which induces the concentration of valence neutron wave functions around the correlating 2$\alpha$.
A linear-chain structure of $^{16}$C is also discussed. 
\end{abstract}


\maketitle

\section{Introduction}\label{introduction}

The linear-chain structure of three $\alpha$ clusters in C isotopes has been a long standing problem, 
for example, as discussed in Refs.~\cite{Morinaga_4N_56,Morinaga_12C_linear_chain_66,vonOertzen_Cisotopes_97,Itagaki_Cisotopes_01}.
Historically, the $0^{+}_{2}$ state in $^{12}$C was suggested to be 
a candidate for the linear-chain structure of three $\alpha$ clusters \cite{Morinaga_4N_56,Morinaga_12C_linear_chain_66}.
However, according to $3\alpha$ cluster model calculations in the 1970s, 
it was pointed out that the $0^{+}_{2}$ state is not a linear-chain state, but a weak coupling state, 
where three $\alpha$ clusters are weakly interacting with each other
\cite{Suzuki_linear_chain_72,Kamimura_12C_77,Uegaki_12C_77,Kamimura_12C_81}.
In recent years, the $0^{+}_{2}$ state attracts an interest 
as the $\alpha$ condensed state in analogy to Bose Einstein condensation in nuclear matter
\cite{Tohsaki_12C_01,Funaki_12C_03,Funaki_12C_05}.
Instead of the $0^{+}_{2}$ state, the $0^{+}_{3}$ state was suggested to have 
a geometric configuration of three $\alpha$ clusters \cite{Kato_kinear-chain_87}. 
In the recent works on $^{12}$C with antisymmetrized molecular dynamics (AMD) 
and fermionic molecular dynamics (FMD) calculations \cite{Neff_12C_04,En'yo_12C_07,Suhara_AMD_10}, 
the $0^{+}_{3}$ state was suggested to have an open triangle structure of three $\alpha$ clusters
and was assigned to the experimental $0^{+}$ state at 10.3 MeV.
However no linear-chain structure of three $\alpha$ clusters nor linear-chain band was found in the $^{12}$C system, 
and therefore, searching for a linear-chain state in C isotopes is still an attractive subject.

Linear-chain structures in C isotopes have been discussed 
in earlier works with 3$\alpha$+$Xn$ cluster models \cite{vonOertzen_Cisotopes_97,Itagaki_Cisotopes_01,Millin_13C_02,vonOertzen_14C_04}
and the Skyrme mean-field model \cite{Maruhn_linear_chain_10}.
In the study by Itagaki {\it et al}. \cite{Itagaki_Cisotopes_01}, 
they assumed a symmetric 3$\alpha$ cluster configuration
with valence neutrons in molecular orbitals.
It was argued that the linear-chain structure of $^{14}$C may be unstable 
because there is no local minimum in the energy curve against the bending mode.

However, to confirm the stability of an excited state, it is essential to take into
account the orthogonality to lower states.
In contrast to the work of Ref.~\cite{Itagaki_Cisotopes_01}, in our previous study on $^{14}$C 
with an AMD method \cite{Suhara_14C_10}, we found the linear-chain structure and its rotational band consisting of 
the $0^{+}_{5}$, $2^{+}_{6}$, and $4^{+}_{6}$ states in $^{14}$C.
In Ref.~\cite{Suhara_14C_10}, we adopted a method 
of $\beta$-$\gamma$ constraint AMD in combination with the generator coordinate method (GCM)
called the $\beta$-$\gamma$ constraint AMD + GCM \cite{Suhara_AMD_10}.
Although the 3$\alpha$ core is not {\it a priori} assumed in the framework, 
various 3$\alpha$+$2n$ structures are obtained in the energy variation. 
As a result of energy variation and GCM calculations, the linear-chain states appear and show $^{10}$Be+$\alpha$ correlation. 
Here, the $^{10}$Be+$\alpha$ correlation means an asymmetric $2\alpha$+$\alpha$ configuration with 
two valence neutrons distributing around the correlating $2\alpha$. 
We note that three $\alpha$ clusters have one geometric configuration, 
and therefore $^{10}$Be+$\alpha$ ($2\alpha$+$\alpha$) correlation does not have a weak coupling feature,  
but a strong coupling feature.
In other words, this $^{10}$Be is not an isolated one, but a polarized configuration 
of the middle $\alpha$ cluster and valence neutrons in the linear-chain structure.

For excited states of $^{14}$C, there exist many experimental works suggesting 
the $^{10}$Be+$\alpha$ cluster structure in the observed excited states above the $^{10}$Be+$\alpha$ threshold
\cite{vonOertzen_14C_04,Milin_14C_04,Soic_CandBisotopes_04,Price_14C_07,Haigh_14C_08}.
We expect that these states might be possible candidates for the linear-chain states.
Then we consider that the $^{10}$Be+$\alpha$ correlation is one of the key features of the linear-chain structure.

Our aim is to confirm the $^{10}$Be+$\alpha$ correlation in the 
linear-chain states of $^{14}$C.
As discussed in Ref.~\cite{Suhara_14C_10}, our previous work indicates that 
the stability of the linear-chain structure in $^{14}$C is owed 
to the valence neutron effects and the orthogonality to lower states.
Since the bending mode is suppressed by the orthogonality in a $^{14}$C system, 
a $3\alpha$ cluster core with a straight line configuration is considered to be rather stable against the bending mode.
Therefore, in the present study, we consider only the linear configuration and 
omit the bending mode to find the origin of the $^{10}$Be+$\alpha$ correlation.
For $^{14}$C, this ansatz is probably valid.
In such a linear system, the $^{10}$Be+$\alpha$ correlation can be described by the asymmetry of 
three $\alpha$ clusters and valence neutron motion. 
Here the first question is 
whether three $\alpha$ clusters favor an asymmetric $2\alpha$+$\alpha$ configuration
which were not incorporated in the work by Itagaki {\it et al.} \cite{Itagaki_Cisotopes_01}. 
The second question is whether the two neutrons localize and form the $^{10}$Be cluster
or if they are moving around all three $\alpha$ clusters as described 
in the molecular orbital picture \cite{vonOertzen_14C_04}. Unfortunately, 
in our method used in the previous work \cite{Suhara_14C_10}, detailed behavior of valence neutrons 
in the molecular orbitals is not sufficiently described because a single particle wave function
is approximated by a Gaussian wave packet.

In the present paper, to investigate the linear-chain structures of $^{14}$C
we extend the molecular orbital model by explicitly incorporating 
asymmetric configurations of the linear-chain structure.
We solve the valence neutron motion in the linear $3\alpha$ system 
by using a linear combination of $p$ orbits around $\alpha$ clusters as done in the molecular orbital model. 
By analyzing the calculated results,
we will confirm the $^{10}$Be+$\alpha$ correlation in the linear-chain structure of $^{14}$C and clarify its origin. 
To see the effect of three $\alpha$ clusters and that of valence neutron motion,  
we compare the results of $^{14}$C with those of $3\alpha$ and $3\alpha$+$n$.
Here, we comment that these 3$\alpha$ and $3\alpha$+$n$ do not necessarily correspond to excited states of
realistic $^{12}$C and $^{13}$C systems.
Although the $0^{+}_{3}$ state of $^{12}$C is suggested to have an open triangle structure,
it is different from the straight-line chain structure assumed in the present analysis. 
Even if a linear-chain structure does not exist in realistic  $^{12}$C and $^{13}$C,
analysis of $3\alpha$ and $3\alpha$+$n$ systems with the linear-chain configuration is useful to
understand the origin of $^{10}$Be+$\alpha$ correlation suggested in the realistic linear-chain states in $^{14}$C.
We also calculate $^{16}$C and discuss systematics of correlations in the linear-chain structure of C isotopes.

This article is organized as follows. 
In Sec.~\ref{framework}, we explain the framework of the generalized molecular orbital model.
The calculated results are shown in Sec.~\ref{results}.
We discuss the motion of valence neutrons and correspondence to the previous work in Sec.~\ref{discussion}. 
Finally, in Sec.~\ref{summary}, we give a summary.

\section{Framework of the generalized molecular orbital model}\label{framework}

In this section, we explain the framework of the generalized molecular orbital model.
In this model, we express the valence neutron motion in the linear-chain structures
by a linear combination of configurations written by the $p$ orbits centered at $\alpha$ clusters.

\subsection{Wave function of the generalized molecular orbital model}

In the present model, the wave function of a carbon isotope with a mass number $A$, $^{A}$C, 
is described by a superposition of basis wave functions
which consist of three $\alpha$ clusters with a linear-chain configuration and 
$X=A-12$ valence neutrons with various configurations:
\begin{equation}
	{\Phi_{^{A}\text{C}}} = \sum_{\bm{c}} f(\bm{c})
		\mathcal{A} \{ \varphi^{\alpha}_{\text{L}}, \varphi^{\alpha}_{\text{M}}, \varphi^{\alpha}_{\text{R}},
		\varphi^{n}_{1} (c_{1}), \cdots, \varphi^{n}_{X} (c_{X}) \},
	\label{wavefunction_generalizedMO}
\end{equation}
where $\mathcal{A}$ is the antisymmetrizing operator and 
$\bm{c}$ represents a set of the configuration for valence neutrons, $\bm{c} \equiv \{c_{1}, \cdots, c_{X} \}$.
The subscripts L, M, and R mean left, middle, and right positions for three $\alpha$ clusters, respectively.
We label the left, middle, and right $\alpha$ clusters $\alpha_\text{L}$, $\alpha_\text{M}$, and $\alpha_\text{R}$. 
The coefficients $f(\bm{c})$ are determined by diagonalization of Hamiltonian and norm matrices as explained later.

\begin{figure}[t]
	\centering
	\includegraphics[width=7.9cm, clip]{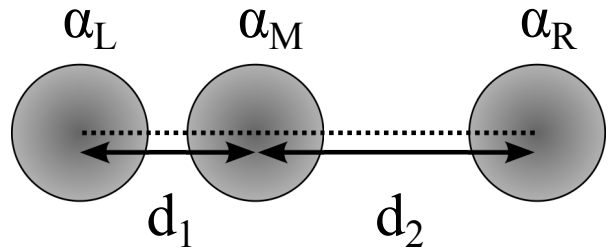}
	\caption{The schematic figure for a linear-chain structure of three $\alpha$ clusters.
	The distances between $\alpha$ clusters L and M, and M and R are set to $d_{1}$ and $d_{2}$, respectively.}
	\label{linear_chain_figure}
\end{figure}

The wave function of $\alpha_{t}$ is given by the $(0s)^{4}$ configuration whose center is shifted at a certain
position $\bm{R}_{t}$, 
\begin{align}
	\varphi^{\alpha}_{t} &= \phi_{\bm{R}_{t}} \phi_{\bm{R}_{t}} \phi_{\bm{R}_{t}} \phi_{\bm{R}_{t}}  
		\chi_{p \uparrow} \chi_{p \downarrow} \chi_{n \uparrow} \chi_{n \downarrow}, \\
	\phi_{\bm{R}_{t}} &= \left( \frac{2\nu}{\pi} \right)^{\frac{3}{4}}
		\exp \left[ - \nu \left( \bm{r} - \frac{\bm{R}_{t}}{\sqrt{\nu}} \right)^{2} \right] 
	\label{single_particle_spatial_generalizedMO},
\end{align}
where $t = \text{L}, \text{M}, \text{R}$. 
A linear-chain structure is expressed by the positions $\bm{R}_{t}$ on the $z$-axis and it is characterized by inter-cluster distances, $d_{1}$ and $d_{2}$:
\begin{align}
	d_{1} &\equiv |\bm{R}_{\text{M}} - \bm{R}_{\text{L}}|, \\
	d_{2} &\equiv |\bm{R}_{\text{R}} - \bm{R}_{\text{M}}|,
\end{align}
as seen in Fig.~\ref{linear_chain_figure}.
Without loss of generality, we can limit the distances $d_{2} \ge d_{1}$.

For spatial wave functions of valence neutrons, we define the $p$ orbits centered at one of
three $\alpha$ clusters,
\begin{align}
	(p_{\pm})_{t} &\equiv \frac{1}{\sqrt{2}} \left((p_{x})_{t} \pm i (p_{y})_{t} \right),\\
	(p_{x})_{t} &= \phi_{\bm{R}_{t} + \delta \bm{e}_{x}} - \phi_{\bm{R}_{t} - \delta \bm{e}_{x}}, \\
	(p_{y})_{t} &= \phi_{\bm{R}_{t} + \delta \bm{e}_{y}} - \phi_{\bm{R}_{t} - \delta \bm{e}_{y}},
\end{align}
where $\delta$ is a small value.
The $(p_{\pm})_{t}$ represents the $p$ orbits around $\alpha_{t}$, and it has
the $z$-component of orbital angular momentum $l_{z}=\pm 1$.
In general, the $k$-th valence neutron wave function $\varphi^{n}_{k} (c_{k})$ in Eq.~\ref{wavefunction_generalizedMO}
is given by a product of $(p_{\pm})_{t}$ and a spin wave function, 
\begin{equation}
\varphi^{n}_{k} (c_{k})=(p_{\pm})_{t}\chi_{n\uparrow} \ \ {\rm or}\ \  (p_{\pm})_{t}\chi_{n\downarrow}.
\end{equation}
$(p_{+})_{t}\chi_{n\uparrow}$ and $(p_{-})_{t}\chi_{n\downarrow}$ are the eigenstates of the $z$-component of 
total angular momentum $j_{z}$ with $j_z\equiv\Omega=+ 3/2$ and $\Omega=- 3/2$, respectively, which are energetically favored 
because of the spin-orbit force, while $(p_{+})_{t}\chi_{n\downarrow}$ and $(p_{-})_{t}\chi_{n\uparrow}$
are the unfavored $\Omega=\pm 1/2$ states.
In the present model, we omit the $p_{z}$ orbits because  
the orbits constructed from the $p_{z}$ orbits have higher nodes than those from $p_{x}$ and $p_{x}$ orbits;
therefore the $p_{z}$ orbits are unfavored.

In the present calculation, we assume the first and second valence neutrons occupy
the favored orbits and the third and fourth valence neutrons occupy the unfavored ones. As a result, 
the valence neutron wave functions can be written with the labels $\Omega=\pm 3/2, \pm 1/2$ and 
$t=\text{L},\text{M},\text{R}$ as, 
\begin{align}
\varphi^{n}_{1} (t_{1}) &= (p_{+})_{t_1} \chi_{n\uparrow}\equiv [+3/2]_{t_1}, \\
\varphi^{n}_{2} (t_{2}) &= (p_{-})_{t_2} \chi_{n\downarrow}\equiv [-3/2]_{t_2}, \\
\varphi^{n}_{3} (t_{3}) &= (p_{+})_{t_3} \chi_{n\uparrow}\equiv [+1/2]_{t_3}, \\
\varphi^{n}_{4} (t_{4}) &= (p_{-})_{t_4} \chi_{n\downarrow}\equiv [-1/2]_{t_4},
\end{align}
and a set of the configuration for valence neutrons, $\bm{c}$, can be explicitly written as
$\bm{c}=\{t_1,\cdots,t_X\}$ with $t_i = \text{L}, \text{M}, \text{R}$. 
It means that the total wave function is a superposed wave function given by the configuration mixing 
for valence neutrons with respect to $t_i = \text{L}, \text{M}, \text{R}$ indicating 
positions of the $p$ orbit centers, 
\begin{equation}
	{\Phi_{^{A}\text{C}}} = \sum_{t_1,\cdots,t_X} f(t_1,\cdots,t_X)
		\mathcal{A} \{ \varphi^{\alpha}_{\text{L}}, \varphi^{\alpha}_{\text{M}}, \varphi^{\alpha}_{\text{R}},
		\varphi^{n}_{1} (t_{1}), \cdots, \varphi^{n}_{X} (t_{X}) \}.
		\label{wavefunction_generalizedMO_2}
\end{equation}
The coefficient $f(t_1,\cdots,t_X)$ is determined by the diagonalization to minimize the energy of the 
state, ${\Phi_{^{A}\text{C}}}$, for a given $3\alpha$ configuration parametrized by $d_1$ and $d_2$.
 
In this model, we introduce an asymmetric correlation of a spatial configuration of $\alpha$ clusters 
by the position of the middle $\alpha$ cluster.
When the middle $\alpha$ cluster shifts to one side considerably, $d_{2} - d_{1} \gg 0$ fm,
a linear three $\alpha$ system has the correlating $2\alpha$ and the isolated $\alpha$ configuration.
This improvement is the most different point from symmetric molecular orbital models. 
Moreover, we incorporate a multineutron correlation by the configuration mixing of valence neutron orbits.

\subsection{Parity and angular momentum projections}

We project the wave functions onto parity and angular momentum eigenstates
by using the parity projection operator $\hat{P}^{\pm}$ and the angular-momentum projection operator $\hat{P}^{J}_{MK}$.
The parity projection operator $\hat{P}^{\pm}$ is defined as 
\begin{equation}
	\hat{P}^{\pm} \equiv \frac{1 \pm \hat{P}}{2},
\end{equation}
where $\hat{P}$ is the parity operator.
The angular-momentum projection operator $\hat{P}^{J}_{MK}$ is defined as
\begin{equation}
	\hat{P}^{J}_{MK} \equiv \frac{2 J + 1}{8 \pi^{2}} \int d \Omega D^{J*}_{MK}(\Omega) \hat{R}(\Omega),
\end{equation}
where $\Omega = (\alpha, \beta, \gamma)$ are the Euler angles, $D^{J}_{MK}(\Omega)$ is the Wigner's $D$ function, 
and $\hat{R}(\Omega) = e^{- i \alpha \hat{J}_z} e^{- i \beta \hat{J}_y} e^{- i \gamma \hat{J}_z}$ is the rotation operator.

We will show energy curves of intrinsic, parity projected, and parity and angular-momentum projected wave functions.

\subsection{Interaction and parameters}

For the effective two-body interactions,
we use the Volkov No.~2 interaction \cite{Volkov_No2_65} as the central force
and the spin-orbit term of the G3RS interaction \cite{G3RS_79} as the $LS$ force.
We take the same interaction parameters as those in Refs.~\cite{Suhara_AMD_10,Suhara_14C_10}, i.e.,
the Majorana exchange parameter $M = 0.6$ ($W = 0.4$), the Bartlett exchange parameter $B = 0.125$,
and the Heisenberg exchange parameter $H = 0.125$ in the central force, and 
$u_{1} = -1600$ MeV and $u_{2} = 1600$ MeV in the $LS$ force.
These parameters are 
the same as those adopted in the studies for $^{9}$Be \cite{Okabe_parameter_79}, and $^{10}$Be \cite{Itagaki_10Be_00}, 
except for a small modification in the strength of the spin-orbit force
to fit the $0^{+}_{1}$ energy of $^{12}$C \cite{Suhara_AMD_10}.

For the width parameter of single-particle Gaussian wave packets in Eq.~\eqref{single_particle_spatial_generalizedMO}, 
we used the value $\nu = 0.235$ fm$^{-2}$, which  is also the same as those in the studies for C isotopes 
\cite{Suhara_AMD_10,Suhara_14C_10,Itagaki_Cisotopes_01,Itagaki_14C_04}.

\subsection{One Slater determinant wave function}\label{subsec:one_slater}

To discuss the correlation between valence neutrons in a linear-chain structure,
which may be incorporated by the diagonalized wave function with the configuration mixing in 
Eq.~\ref{wavefunction_generalizedMO_2}, 
we also introduce one Slater determinant wave function having no explicit two-body correlation in principle, 
and compare the results.

In the framework of one Slater determinant wave function, 
each single particle wave functions for a valence neutron is given by a linear combination of $[\Omega]_t$ 
orbits with respect to $t=\text{L}, \text{M}, \text{R}$  
instead of the configuration mixing in Eq.~\ref{wavefunction_generalizedMO_2}:
\begin{equation}\label{eq:one_slater}
	\Phi^{\text{OneSlater}}_{^{A}\text{C}} = 
		\mathcal{A} \{ \varphi^{\alpha}_{\text{L}}, \varphi^{\alpha}_{\text{M}}, \varphi^{\alpha}_{\text{R}},
		\pi^{-}(\Omega_{1}), \cdots, \pi^{-}(\Omega_{X}) \}.
\end{equation}
where
\begin{equation}
	\pi^{-}(\Omega) = \sum_{t=\text{L}, \text{M}, \text{R}} C_{\Omega,t} [\Omega]_{t}. 
\end{equation}
$\Omega_1 = +3/2$, $\Omega_2 = -3/2$, $\Omega_3 = +1/2$, and $\Omega_4 = -1/2$ are chosen
for the first, second, third and fourth valence neutrons as well as Eq.~\ref{wavefunction_generalizedMO_2}
and $C_{\Omega,t}$ are the coefficients for the orbits $[\Omega]_{t}$.
$\pi^{-}(\Omega)$ means the character of the orbit: 
the $\pi$ orbit, negative parity, and $z$ component of angular momentum $\Omega$. 
In the limit of $C_{\Omega,\text{L}} = C_{\Omega,\text{M}} = C_{\Omega,\text{R}} =1$ and $d_{1} = d_{2}$, 
these $\pi^{-}(\Omega)$ orbits coincide with the $\pi$ orbits of a usual molecular orbital model.

In the present study, the coefficients $C_{\Omega,t}$ are determined by a variational calculation.
In a variational calculation, we apply the following conditions for the coefficients,
\begin{align}
	C_{+3/2,t} &= C_{-3/2,t}, \\
	C_{+1/2,t} &= C_{-1/2,t}.
\end{align}
The model space spanned by this one Slater determinant wave function is
a subspace of the generalized molecular orbital model.

We also introduce three kinds of test wave functions with fixed coefficients
which are expected from naive pictures.
We name three orbits `MO', `PO', and `AO'. 
In `MO', the coefficients are chosen as
\begin{equation}
	C_{\Omega,\text{L}} = C_{\Omega,\text{M}} = C_{\Omega,\text{R}} = 1,
\end{equation}
which indicates the molecular orbital limit where valence neutrons spread over the whole of three $\alpha$ clusters. 
`AO' corresponds to the atomic orbital limit with valence neutrons localizing around the middle $\alpha$ cluster:
\begin{align}
	C_{\Omega,\text{M}} &= 1,\\
	C_{\Omega,\text{L}} &= C_{\Omega,\text{R}} = 0.
\end{align}
`PO' means the partial orbit in which valence neutrons distribute around the correlating $2\alpha$ as
\begin{align}
	C_{\Omega,\text{L}} &= C_{\Omega,\text{M}} =1, \\ 
	C_{\Omega,\text{R}} &= 0.
\end{align}
This distribution describes a naively expected Be correlation. 
However, please note that realistic Be correlation
in the linear-chain state of $^{14}$C is different from this 'PO' configuration as shown later.
For example, the $\pi^{-}(+3/2)$ orbits for `MO', `PO', and `AO' are written as follows:
\begin{eqnarray}
	\text{`MO'}&: & \pi^{-}(+3/2) = 
		[+3/2]_{\text{L}} + [+3/2]_{\text{M}} + [+3/2]_{\text{R}} , \\
	\text{`PO'}&: & \pi^{-}(+3/2) = 
    	[+3/2]_{\text{L}} + [+3/2]_{\text{M}} , \\
	\text{`AO'}&: & \pi^{-}(+3/2) = [+3/2]_{\text{M}}. 
\end{eqnarray}
Neutrons in `MO' orbits gain the kinetic energy because valence neutrons distribute 
in a wide region along the $z$-axis, while those in `AO' orbits gain maximum potential energy, as shown later.

\section{Results}\label{results}

We applied the generalized molecular orbital model to C isotopes: $^{14}$C, $^{12}$C, $^{13}$C, and $^{16}$C. 
In this section, we show the calculated results.

\begin{figure}
	\centering
	\begin{tabular}{c}
	\includegraphics[width=8.6cm]{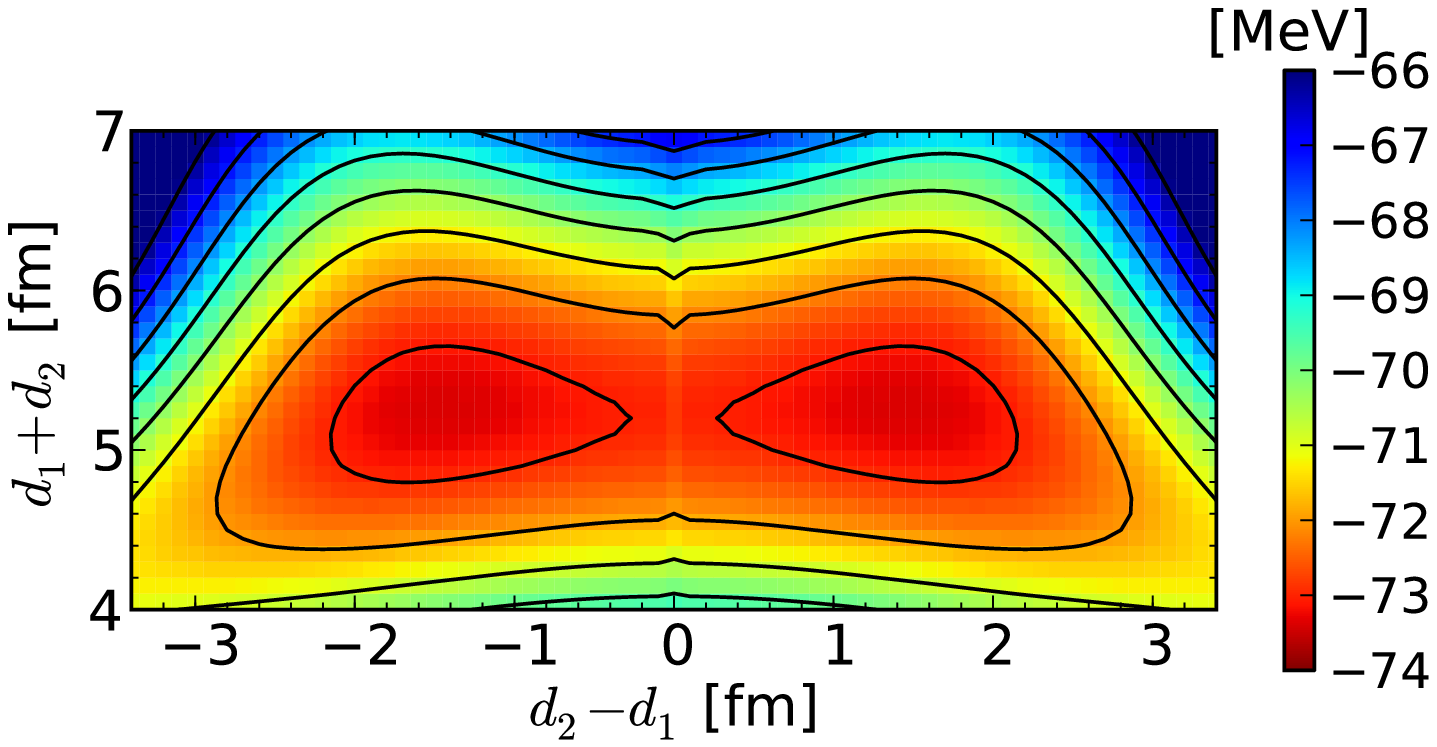} \\
	\includegraphics[width=8.6cm]{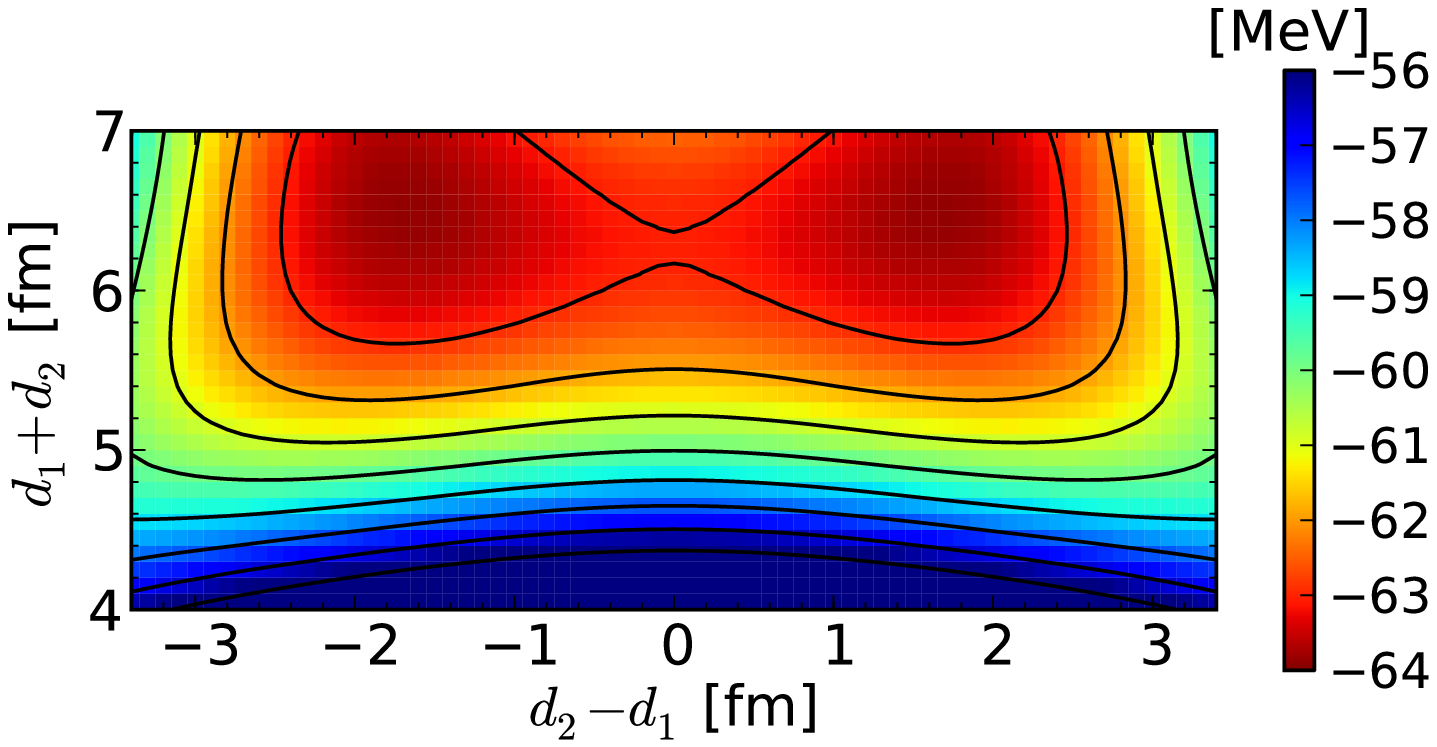} \\
	\includegraphics[width=8.6cm]{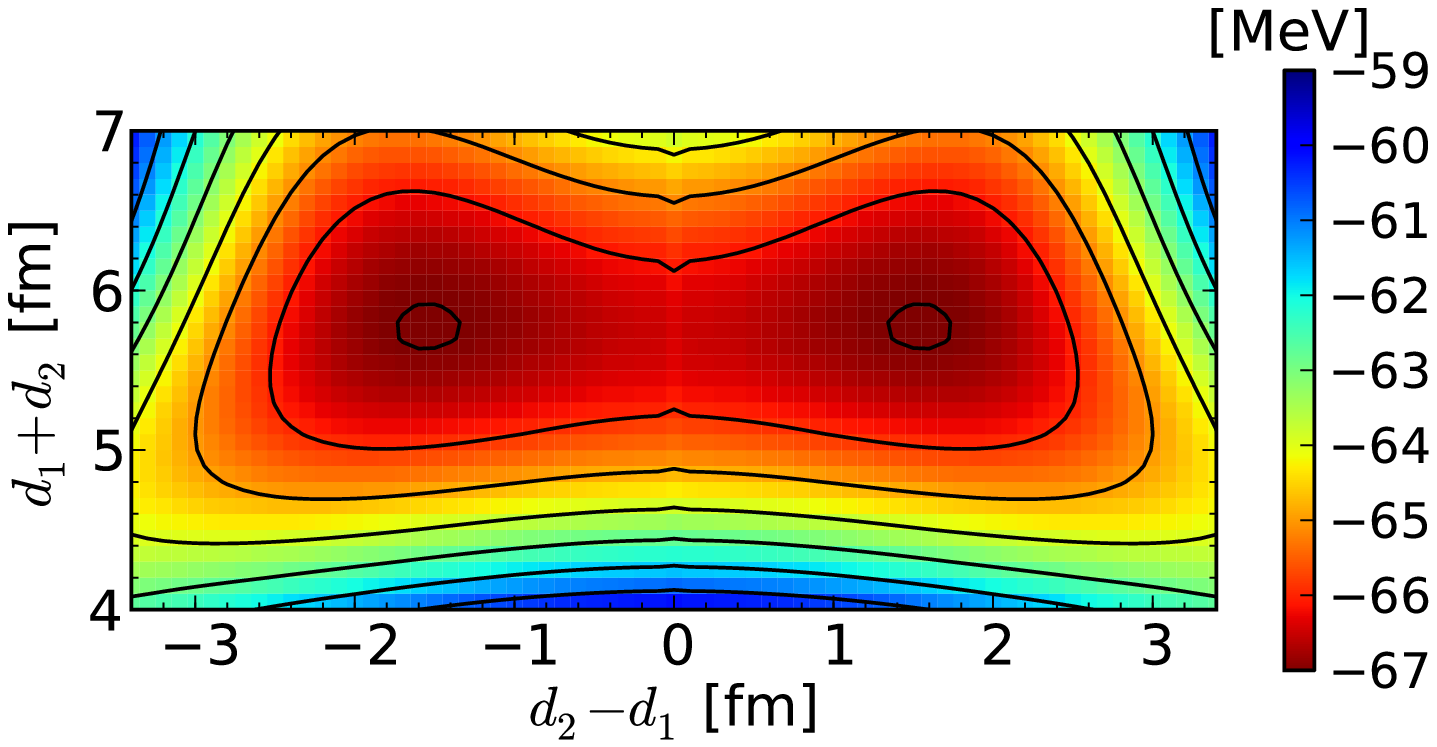} 
	\end{tabular}
	\caption{(Color online) The energy surfaces of positive-parity states in $^{14}$C,
	positive-parity states in $^{12}$C, and negative-parity states in $^{13}$C. 
	The $\alpha$-$\alpha$ distances, $d_{1}$ and $d_{2}$, are taken as described in Fig.~\ref{linear_chain_figure}.}
	\label{14C_energy_surface}
\end{figure}

The energy surfaces of positive-parity states in $^{14}$C, 
positive-parity states in $^{12}$C, and negative-parity states in $^{13}$C 
are shown as functions of $d_{2} - d_{1}$ and $d_{1} + d_{2}$ in Fig.~\ref{14C_energy_surface}. 
$d_{1} + d_{2}$ means the total length of the linear chain, while $d_{2} - d_{1}$ is 
the difference of the intervals between $\alpha$ clusters and it indicates asymmetry
of the $3\alpha$ configuration (see Fig.~\ref{linear_chain_figure}). 
Namely, linear three $\alpha$ clusters have symmetric structure for $d_{2} - d_{1} = 0$ fm, 
and the asymmetric structure develops as the $d_{2} - d_{1}$ value increases. 
At the minimum point of the energy surface of $^{14}$C,
the linear-chain length $d_{1} + d_{2}$ and the difference
$d_{2} - d_{1}$ are 5.2 fm and 1.5 fm, respectively. 
The existence of the energy minimum at the finite $d_{2} - d_{1}$ value indicates $^{10}$Be+$\alpha$ correlation in $^{14}$C,
though this minimum is shallow.
At the minimum points of the energy surfaces of $^{12}$C and $^{13}$C,
the differences $d_{2} - d_{1}$ are 1.6 fm and 1.5 fm, respectively, which are similar to the $^{14}$C case.
However, the linear-chain lengths $d_{1} + d_{2}$ are 6.5 fm and 5.8 fm, respectively, which are different from the $^{14}$C case.
More valence neutrons shorten the linear-chain length but do not change the difference of the intervals between $\alpha$ clusters.
Although the locations of energy minima are different depending on the number of valence neutrons,
all the energy surfaces indicate Be+$\alpha$ correlation in the linear-chain structures of C isotopes.
Here, we stress again that the aim of calculating the $^{12}$C and $^{13}$C systems 
with linear $3\alpha$ and $3\alpha$+n structures is to clarify the origin of the $^{10}$Be+$\alpha$ correlation in $^{14}$C.
In the following, we fixed the length $d_{1} + d_{2}$ to be 5.2 fm and show the energy curves of the linear-chain structure
as a function of the difference $d_{2} - d_{1}$
because the energy surface of $^{14}$C is rather soft over the wide range of the difference $d_{2} - d_{1}$
but is steep in the linear-chain length $d_{1} + d_{2}$ around the $d_{1} + d_{2} = 5.2$ fm line.

\begin{figure}[t]
	\centering
	\begin{tabular}{c}
	\includegraphics[width=7.9cm]{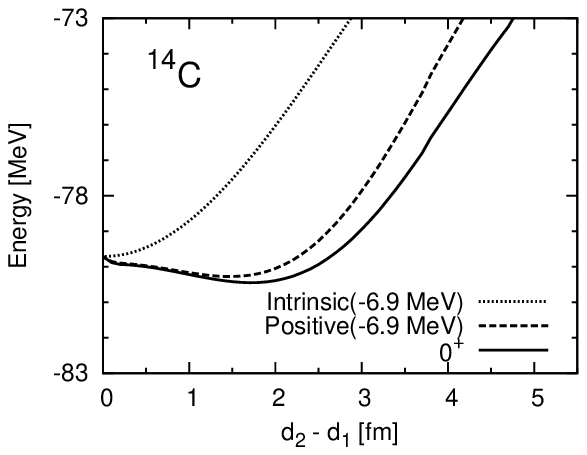} \\
	\includegraphics[width=7.9cm]{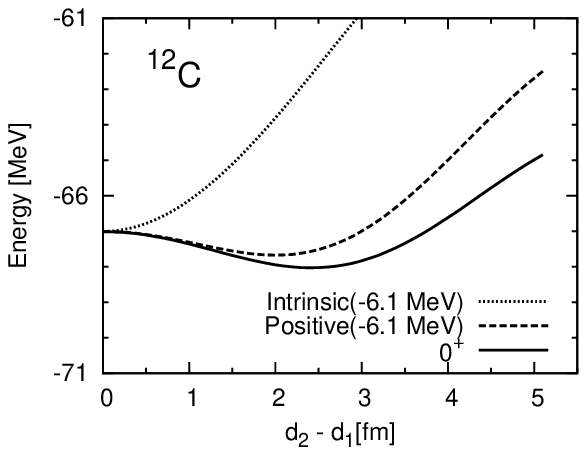} \\
	\includegraphics[width=7.9cm]{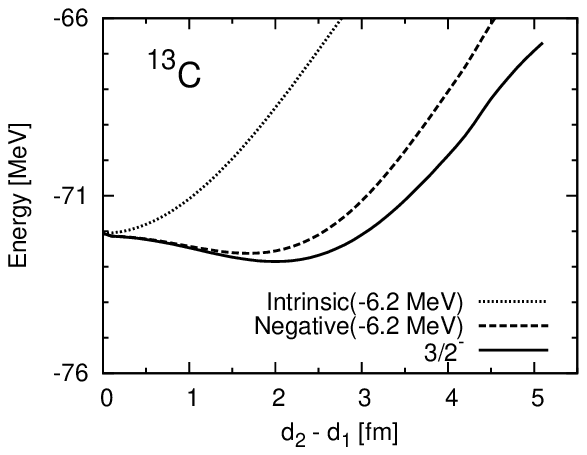} 
	\end{tabular}
	\caption{Energy curves of $^{14}$C, $^{12}$C, and $^{13}$C
		with the linear-chain structure whose length $d_{1} + d_{2}$  is fixed to be 5.2 fm,
		as a function of the difference of the $\alpha$-cluster intervals, $d_{2} - d_{1}$.
		Each panel shows the energy curves for the intrinsic, parity projected, 
		and angular-momentum projected wave functions of the generalized molecular orbital model.
		The intrinsic and parity projected energy curves are shifted to agree with 
		the angular-momentum projected energy curves at $d_{1} - d_{2} = 0$ fm.}
	\label{energy_alpha_motion_5.2fm}
\end{figure}

In Fig.~\ref{energy_alpha_motion_5.2fm}, we show the energy curves of $^{14}$C, $^{12}$C, and $^{13}$C
with the linear-chain structure. 
In each panel, the energy curves for the intrinsic, parity projected, 
and angular-momentum projected wave functions of the generalized molecular orbital model are shown.
To discuss asymmetry of the linear-chain structure, we shift the energy curves for the intrinsic and 
parity projected wave functions of $^{14}$C, $^{12}$C, and $^{13}$C by $-6.9$ MeV, $-6.1$ MeV, 
and $-6.2$ MeV, respectively, to adjust the energies to the angular-momentum projected energies at
$d_{2} - d_{1} = 0$ fm.

At first, we explain the results for $^{14}$C shown in the top panel of Fig.~\ref{energy_alpha_motion_5.2fm}.
The intrinsic energy curve has the energy minimum at the symmetric configuration, $d_{2} - d_{1} = 0$ fm,
while the positive-parity and the $0^{+}$ energy curves have energy minima at asymmetric configurations, $d_{2} - d_{1} > 0$ fm.
This difference indicates that the many body correlation incorporated by the parity projection 
gain the energy with the asymmetric configuration of three $\alpha$ clusters resulting in the $2\alpha$+$\alpha$ correlation.
The kinetic energy gain by the parity projection, which spreads the center of mass motion of $\alpha$ clusters,
becomes larger than the interaction energy loss in the asymmetric configuration at the energy minimum.
The depth of the $0^{+}$ minimum, 0.73 MeV, is deeper than that of the positive-parity minimum, 0.56 MeV.
Moreover the position of the $0^{+}$ minimum, $d_{2} - d_{1} = 1.7$ fm, 
is more asymmetric than that of the positive-parity minimum, $d_{2} - d_{1} = 1.5$ fm.
That is, many body correlation incorporated by the angular-momentum projection
enhances the the asymmetric correlation in the present calculation.

Next, we discuss the results for $^{12}$C (the middle panel of Fig.~\ref{energy_alpha_motion_5.2fm}) 
to extract the contribution of three $\alpha$ clusters 
without valence neutrons to the asymmetric correlation.
The behaviors of the energy curves are qualitatively similar to that of $^{14}$C.
The positive-parity and $0^{+}$ energy curves  have minima at asymmetric configurations,
while the intrinsic energy curve has the minimum at $d_{2} - d_{1} = 0$ fm of the symmetric configuration.
The depth of the $0^{+}$ minimum, 1.02 MeV, is deeper than that of the positive-parity minimum, 0.67 MeV.
Again, the degree of the asymmetry of the $0^{+}$ energy curve ($d_{2} - d_{1} = 2.4$ fm) is larger than 
that of the positive-parity energy curve ($d_{2} - d_{1} = 2.0$ fm).
Interestingly, even if valence neutrons are absent, a 3$\alpha$ system prefers an asymmetric configuration 
resulting in the $2\alpha$ correlation after the parity projection.
That is, linear three $\alpha$ clusters have the $^{8}$Be+$\alpha$ correlation, 
and $^{10}$Be+$\alpha$ correlation in the linear chain states of $^{14}$C originates in this nature
of $3\alpha$ systems.
Compared with the parity projected and $0^{+}$ energy curves for $^{12}$C with those for $^{14}$C, 
it is found that energy curves in the large $d_{2} - d_{1}$ region are less steep, and 
the energy minimum points are deeper and more asymmetric than those for $^{14}$C.
The former is easily understood from the viewpoint of Be-$\alpha$ inter-cluster potentials.
That is, the potential energy between $^{10}$Be and $\alpha$ clusters is deeper than 
that between $^{8}$Be and $\alpha$ clusters because of the additional two neutrons.

From the results for $^{13}$C shown in the bottom panel of Fig.~\ref{energy_alpha_motion_5.2fm}, 
we can see the effect of a valence neutron in the linear-chain structure.
The behaviors of the energy curves are qualitatively analogous to those of $^{14}$C and $^{12}$C.
Although the intrinsic energy curve has the minimum at the symmetric configuration, 
the negative-parity and the $3/2^{-}$ energy curves have minima at asymmetric configurations.
The depth of the $3/2^{-}$ minimum, 0.79 MeV, is deeper than that of the negative-parity minimum, 0.57 MeV.
The degree of the asymmetry of the $3/2^{-}$ energy surface ($d_{2} - d_{1} = 2.0$ fm) is larger than 
that of the negative-parity energy surface ($d_{2} - d_{1} = 1.7$ fm).

In the systematics of the minima of the angular momentum projected energy curves for $^{12}$C, $^{13}$C, and $^{14}$C, 
it is found that the asymmetry $d_{2} - d_{1}$ is the largest in the linear $3\alpha$ structure without valence neutrons. 
As a valence neutron is added to the linear $3\alpha$ system one by one, 
the depth of minima becomes shallow and the asymmetry $d_{2} - d_{1}$ reduces slightly. 
Therefore  valence neutrons seem to somehow reduce the asymmetry of the linear $3\alpha$ structure. 
However, as we show in the following, valence neutrons gather around the middle of linear three $\alpha$ clusters
and play an important role to enhance the Be+$\alpha$ correlations.

\begin{figure}[t]
	\centering
	\begin{tabular}{c}
	\includegraphics[width=7.9cm]{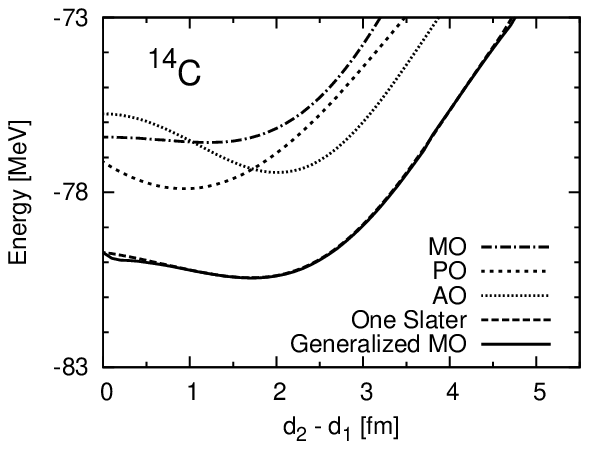} \\
	\includegraphics[width=7.9cm]{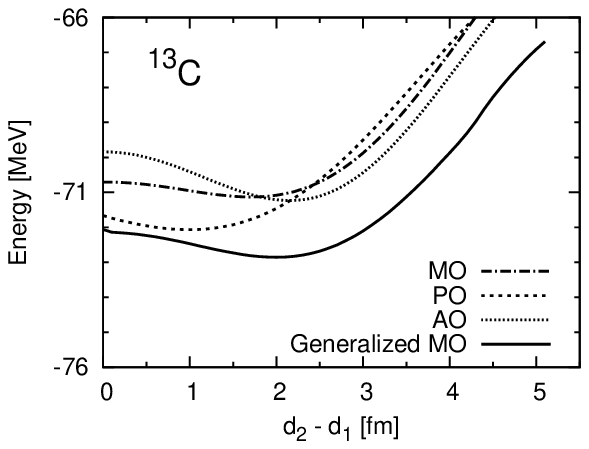}
	\end{tabular}
	\caption{Energy curves of $^{14}$C and $^{13}$C	with the linear-chain structure whose length is fixed to be 5.2 fm. 
		Energies are plotted as a function of the difference of the $\alpha$-cluster intervals, $d_{2} - d_{1}$.
		The top and bottom panels show the $^{14}$C($0^{+}$) and $^{13}$C($3/2^{-}$) energy curves for the `MO', `PO', `AO', 
		and one Slater determinant wave functions compared with the curve for  
		the generalized molecular orbital model wave function.}
	\label{energy_neutron_motion_5.2fm}
\end{figure}

We discuss the motion of valence neutrons and their effect in the linear $3\alpha$ structure by comparing the results of 
the generalized molecular orbital model with those calculated by using simpler wave functions, 
`MO', `PO', `AO', and one Slater determinant defined in Sec. \ref{subsec:one_slater}.
In Fig.~\ref{energy_neutron_motion_5.2fm}, we show the angular-momentum projected energy curves of $^{14}$C and $^{13}$C
as a function of the difference $d_{2} - d_{1}$ of the $\alpha$-cluster intervals.
The top panel shows the $^{14}$C($0^{+}$) energy curves for the `MO', `PO', `AO', and one Slater determinant wave functions 
compared with the generalized molecular orbital ones, and the bottom panel shows the $^{13}$C($3/2^{-}$) energy curves.
It should be noted that one Slater determinant wave function is equivalent to 
the generalized molecular orbital wave functions in $^{13}$C, 
because two models are coincident with each other in the case of a single valence neutron. 
This is not necessarily the case in $^{14}$C because the generalized molecular orbital wave functions contain 
two neutron correlations due to the configuration mixing, while one Slater determinant does not have two neutron correlations.
 
In spite of the difference of the model space, the obtained one Slater determinant
wave functions for $^{14}$C give the same $0^+$ energy curve as that of the generalized molecular orbital model
in all regions of $d_{2} - d_{1}$. Actually, the obtained $^{14}$C wave functions of these two models 
are found to be equivalent to each other with about 99\% overlap.
This means that the two-body correlation between valence neutrons is not essential 
in the linear chain structure of $^{14}$C
but the intrinsic wave functions can be described by a simple product of single-particle wave functions as written by 
one Slater determinant in Eq. \ref{eq:one_slater}. 
Next, we describe the energy curves obtained by using three kinds of test wave functions, `MO', `PO', and `AO'.
Although the energy curves are higher than those of the generalized molecular orbital model 
and one Slater determinant, all of these wave functions show soft energy curves against the
asymmetry $d_{2} - d_{1}$ similarly to the case of the generalized molecular orbital model.
The energy minima of `PO' and `AO' reproduce an asymmetric configuration.
In particular, the degree of the asymmetry of `AO', $d_{2} - d_{1} = 1.9$ fm, is in good agreement
with that of the generalized molecular orbital model,
though the depth of the `AO' energy minimum is larger. 

We discuss valence neutron wave functions in the one Slater determinant wave function of $^{14}$C, 
which almost coincides with that obtained by the generalized molecular orbital model.
As mentioned before, the coefficients $C_{3/2,t}$ ($t=\text{L}, \text{M}, \text{R}$) in Eq. \ref{eq:one_slater}
are optimized for each $d_{2} - d_{1}$.
At the energy minimum, $d_{2} - d_{1} = 1.7$ fm, the ratio of the coefficients is 
\begin{equation}
	C_{3/2,\text{L}} : C_{3/2,\text{M}} : C_{3/2,\text{R}} = 1.0 : 4.3 : 0.6.
\end{equation}
The largest coefficient $C_{3/2,\text{M}}$ indicates that valence neutrons gather around the middle $\alpha$ cluster($\alpha_\text{M}$). 
Since the $\alpha_\text{M}$ is shifted to one side,
the concentration of the amplitude at the $\alpha_\text{M}$ means that
valence neutron distribution leans toward the narrow side of two of three $\alpha$ clusters.
That is, the asymmetric configuration of linear three $\alpha$ clusters 
and valence neutrons following the middle $\alpha$ cluster make $^{10}$Be+$\alpha$ correlation, 
which is consistent with the result in Ref.~\cite{Suhara_14C_10}.
For more quantitative discussion of the asymmetry, 
the reduction of the positive parity component in the intrinsic wave function can be a good measure.
At the $d_{2} - d_{1} = 1.7$ fm corresponding to the energy minimum of linear-chain states in $^{14}$C, 
the magnitude of the positive parity component in the linear 3$\alpha$ system without valence neutrons is 89\%, 
while that with two valence neutrons for the whole $^{14}$C system is found to be 82\%.
The reduction of the positive parity component, i.e., the enhancement of the asymmetry by valence neutrons 
owes to the concentration of the neutron distribution at the middle $\alpha$ cluster. 
Incidentally, 82\% for $^{14}$C at the energy minimum, $d_{2} - d_{1} = 1.7$ fm, 
is very close to the magnitudes of the positive parity component, 81\%, 
in the linear-chain structure of $^{12}$C at the energy minimum, $d_{2} - d_{1} = 2.4$ fm.
In the next section, we show that this localization of valence neutrons can be understood by a mean field potential 
from three $\alpha$ clusters with the linear-chain structure.

In $^{13}$C, the situation is quite similar to $^{14}$C.
Although the energy curves of `MO', `PO', and `AO' are higher than that of the generalized molecular orbital model, 
they give energy minima at asymmetric configurations.
As well as $^{14}$C, the degree of the asymmetry of `AO' at the energy minimum 
agrees well with that of the generalized molecular orbital model.
Here, we also analyze the distribution of the valence neutron in $^{13}$C.
At the energy minimum, the ratio of the coefficients $C_{3/2,t}$ ($j=\text{L}, \text{M}, \text{R}$) is
\begin{equation}
	C_{3/2,\text{L}} : C_{3/2,\text{M}} : C_{3/2,\text{R}} = 1.0 : 4.4 : 0.8.
\end{equation}
As well as $^{14}$C, the valence neutron gathers around the middle $\alpha$ cluster and 
the $^{9}$Be+$\alpha$ correlation is found in $^{13}$C. 

Surprisingly, the ratio of coefficients and the behaviors of energy curves in $^{14}$C are almost the 
same as those in $^{13}$C. 
This is consistent with the fact that the linear $^{14}$C can be described by one Slater determinant wave function
and valence neutrons have no explicit dineutron correlation.
In this analysis, we find that valence neutrons in $^{14}$C are moving rather independently 
in the mean field constructed from the linear three $\alpha$ clusters. In other words, 
a valence neutron in $^{14}$C is not affected by another valence neutron.

\begin{figure}[t]
	\centering
	\includegraphics[width=7.9cm]{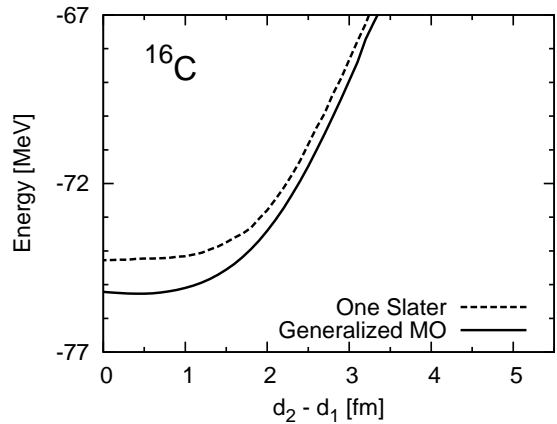}
	\caption{The positive-parity energy curves of $^{16}$C using `MO', `PO', `AO', 
		one Slater determinant, and the generalized molecular orbital model
		with the linear-chain structure whose length is 5.2 fm.}
	\label{16C_energy_+_5.2fm}
\end{figure}

Finally, we show the energy curves of $^{16}$C to study the systematics of correlations 
in linear chain structures of C isotopes.
For $^{16}$C, we do not calculate the $0^{+}$ energy curve because of the computational cost.
However, we have already seen in the results of $^{12,13,14}$C that 
the $0^{+}$ energy curves are essentially the same as those of the positive-parity energy curves. 
Therefore, in this paper, we discuss the positive-parity energy curves of $^{16}$C. 
In Fig.~\ref{16C_energy_+_5.2fm}, we show the positive-parity energy curves of one Slater determinant 
and the generalized molecular orbital model wave functions.
The behaviors of the energy curves are different from those of other C isotopes.
The most striking feature is that the positive-parity energy curve has 
no energy minimum at an asymmetric configuration. Although it is soft around $d_{2} - d_{1} = 0$ fm, 
an asymmetric correlation seems to disappear in $^{16}$C.
Moreover, in $^{16}$C, the energy curve of the generalized molecular orbital model is lower by 1 MeV 
than that of one Slater determinant wave function in the $d_{2} - d_{1} \le 1$ fm region. 
This means the configuration mixing in the generalized molecular orbital model gains some 
correlation energy due to multi-neutron correlation in the linear-chain structure of $^{16}$C.
Comparing the wave function at the symmetric configuration $d_{2} - d_{1} = 0$ fm
between the one Slater determinant model and the generalized molecular orbital model, 
the overlap with each other is found to be 95\%, which is still large, 
but smaller than the 99\% overlap for the case of $^{14}$C. 

The ratios of the coefficients $C_{\Omega,t}$ ($\Omega = 3/2, 1/2$, $t=\text{L}, \text{M}, \text{R}$) in the 
one Slater determinant wave function are
\begin{align}
	C_{3/2,\text{L}} : C_{3/2,\text{M}} : C_{3/2,\text{R}} = 1.0 : 4.9 : 0.5, \\
	C_{1/2,\text{L}} : C_{1/2,\text{M}} : C_{1/2,\text{R}} = 1.0 : 5.0 : 0.7.
\end{align}
As same as $^{14}$C and $^{13}$C, the valence neutrons gather around the middle of the linear $\alpha$ clusters.
However, in $^{16}$C, the $3\alpha$ core does not show the $^{12}$Be+$\alpha$ correlation but it is symmetric
at the energy minimum. 
Furthermore, the valence neutrons also have the symmetric configuration because they distribute around 
the middle $\alpha$ cluster, which exists just at the center of linear-chain structure. 
This result indicates a linear-chain structure in $^{16}$C has a symmetric configuration.

The question to be answered in the future is whether or not the linear-chain structure is stable in excited states of $^{16}$C. 
Firstly, the formation of three $\alpha$ clusters should be checked 
by such a framework free from model assumptions of clusters as the AMD method 
\cite{En'yo_PTP_95,En'yo_AMD_95,En'yo_sup_01,En'yo_AMD_03}. 
Secondly, because the orthogonality to low-lying states is essential for the stabilization 
of linear-chain states as found in $^{14}$C \cite{Suhara_14C_10}, 
it is necessary to investigate the structure of $^{16}$C covering low-lying states to high-lying states systematically.

\section{Discussion}\label{discussion}

\begin{figure}[t]
	\centering
	\includegraphics[width=7.9cm, bb = 123 275 453 518, clip]{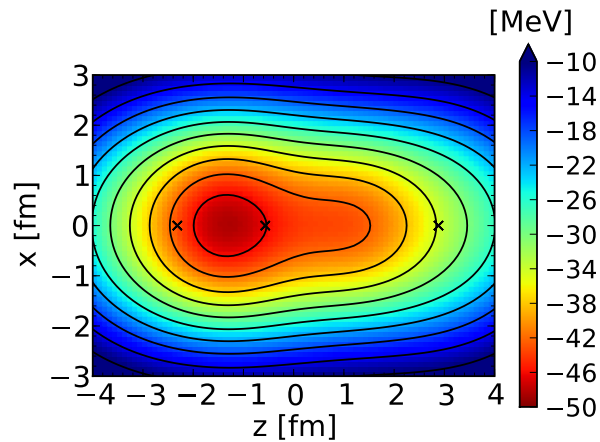}
	\caption{(Color online) The single-folding potential of three $\alpha$ clusters with the linear-chain structure.
	The crosses show the positions of three $\alpha$ clusters.}
	\label{12C_folding_potential}
\end{figure}

In this section, we give further discussion of the behavior of valence neutrons 
in the linear-chain structure. 
As already mentioned, valence neutrons tend to distribute around the middle $\alpha$ cluster. 
It seems in contrast to a simple molecular orbital picture,
where valence neutrons are considered to distribute widely surrounding all $\alpha$ clusters.  
The tendency of the concentration around the middle $\alpha$ is understood by a mean-field potential 
formed by linear three $\alpha$ clusters as follows. 
To see the effect of the mean-field potential for a valence neutron
from linear three $\alpha$ clusters
we calculate the single-folding potential using the Volkov No.~2 force,
\begin{equation}
	U(\bm{r}) = \int d\bm{r^\prime} \rho(\bm{r^\prime}) V^{\text{central}}(\bm{r} - \bm{r^{\prime}}),
	\label{single_folding_potential}
\end{equation}
of three $\alpha$ clusters with the linear-chain structure.
Here $\rho(\bm{r^\prime})$ is the density distribution of linear three $\alpha$ clusters.
In Fig.~\ref{12C_folding_potential}, we show
the contour plot of the folding potential on the $xz$ plane.
Valence neutrons feel this potential in linear three $\alpha$ clusters. 
As clearly seen, the folding potential is deepest in the narrow side of two of three $\alpha$ clusters
and becomes shallow around the isolated $\alpha$ cluster;
therefore, valence neutrons tend to gather around the narrow side to gain potential energy.
Around the narrow side, valence neutrons feel attractions from the two $\alpha$ clusters 
and gain larger potential energy,
while around the isolated $\alpha$ cluster, valence neutrons feel an attraction only from one $\alpha$ cluster 
and gain less potential energy.

\begin{figure}[t]
	\centering
	\includegraphics[width=7.9cm]{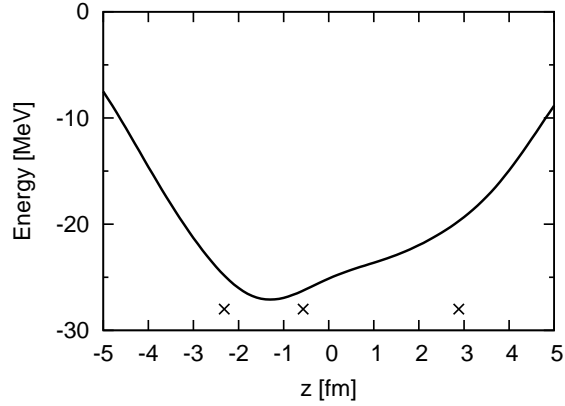}
	\caption{The folding potential of three $\alpha$ clusters with the linear-chain structure for $x = 2.0$ fm.
	The crosses show the positions of three $\alpha$ clusters}
	\label{12C_folding_potential_x=2.0fm}
\end{figure}

Here, we note that valence neutrons do not distribute the deepest region of the potential
because of Pauli blocking with $\alpha$ clusters. In fact, neutron wave functions in $p$ orbits, $(p_\pm)_t$
$(t = \text{L}, \text{M}, \text{R})$, vanish on the $x=y=0$ axis. 
In the present calculations, the root mean square distance $\sqrt{\braket{\rho^{2}}}$ from 
the $x=y=0$ axis for a $p$ orbit 
with the width parameter $\nu = 0.235$ fm$^{-2}$ is
\begin{equation}
	\sqrt{\braket{\rho^{2}}} = \sqrt{\frac{1}{\nu}} = 2.06 \text{ fm}.
\end{equation}
Roughly speaking, valence neutrons in $p$ orbits feel the potential around  
the $\sqrt{\braket{\rho^{2}}} = 2.06 \text{ fm}$ surface of the linear three $\alpha$ structure.
The folding potential along the $x = 2.0$ fm line is shown in Fig.~\ref{12C_folding_potential_x=2.0fm}. 
The behavior of this energy curve is consistent with that seen in Fig.~\ref{12C_folding_potential}. 
That is, the folding potential is deepest in the narrow side of two of three $\alpha$ clusters
and becomes shallow around the isolated $\alpha$ cluster. 
The mean field constructed from linear three $\alpha$ clusters induces the concentration of valence neutron
wave functions around the middle $\alpha$ cluster.

\begin{figure}[t]
	\centering
	\includegraphics[width=7.9cm]{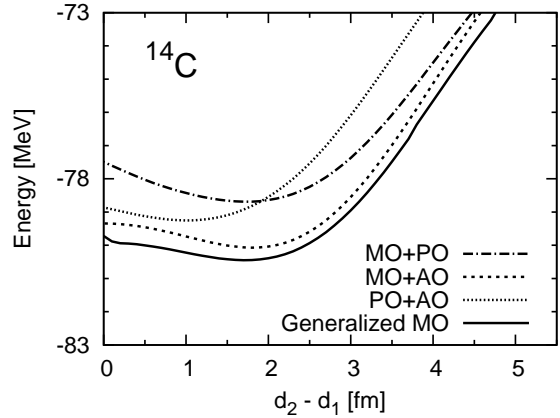}
	\caption{$0^{+}$ energy surfaces of $^{14}$C with the linear-chain structure 
		against the difference between the distances of $\alpha$ clusters.
		The energy surface for the mixing configurations of `MO', `PO', and `AO' are illustrated 
		as well as that of the generalized molecular orbital model.}
	\label{14C_mixing_0+_5.2fm}
\end{figure}

As shown above, the mean-field potential induces the concentration of valence neutron
wave functions around the middle $\alpha$ cluster, while the localization of the wave functions loses kinetic energy. 
As a result of the balance of potential energy gain and kinetic energy loss,
valence neutron wave functions are optimized. 
Here, we describe features of three kinds of test wave functions, `MO', `PO', and `AO', and 
discuss valence neutron motion from a viewpoint of kinetic and potential energies. 
In `MO', valence neutrons spread over the whole of three $\alpha$ clusters.
`PO' describes a naively expected Be correlation with valence neutrons distributing 
around two of the three $\alpha$ clusters.
`AO' means valence neutrons around the middle $\alpha$ cluster.
`MO' gains larger kinetic energy because of the spread of valence neutron wave functions, 
while it gains less potential energy because of no concentration around the middle $\alpha$ cluster.
On the other hand, `AO' gains larger potential energy and looses kinetic energy because of the localization.
The failure to reproduce the results of the generalized molecular orbital model by the `PO' model
indicates that the Be correlation in linear three $\alpha$ clusters is
different from the naively expected Be correlation which is incorporated in the `PO' wave function.
In the obtained results of the generalized molecular orbital model,
both the kinetic energy gain and potential energy gain play important roles.
To demonstrate this feature, we show the $0^{+}$ energy curves of $^{14}$C calculated 
by mixing two of the three `MO', `PO', and `AO' configurations in comparison with the results of 
the generalized molecular orbital model in Fig.~\ref{14C_mixing_0+_5.2fm}.
The point is that the `MO+AO' curve is very close to that of the generalized molecular orbital model.
This indicates that the linear-chain state in $^{14}$C can be interpreted as an intermediate state of `MO' and `AO'
to gain both of the kinetic energy and potential energy.
On the other hand, other sets of mixing, `MO+PO' and `PO+AO' fail to reproduce the results of the generalized molecular orbital model
and `MO+AO' mixing.

Next, we discuss the relation between the present results and our previous results with AMD \cite{Suhara_14C_10}.
In the present results, the minimum point of the $0^{+}$ energy surface ($d_{2} - d_{1} = 1.7$)
has $^{10}$Be+$\alpha$ correlation and 80.4 MeV binding energy.
In the AMD results, the main component of the $0^{+}_{5}$ linear-chain state also has $^{10}$Be+$\alpha$ correlation 
and 80.6 MeV binding energy.
The binding energies are almost the same and the valence neutron behaviors are qualitatively similar. 
Moreover, the overlap between these wave functions is significant as 81\%.
Therefore, we can conclude that the present results correspond to the AMD results.
This good correspondence between the present and AMD results indicates that 
the previous AMD calculation properly describes the valence neutron configurations 
in the linear-chain structure of the $0^{+}_{5}$ in $^{14}$C.
In Ref.~\cite{Suhara_14C_10}, we expected that the linear-chain states (rotational band consisting of
the $0^{+}_{5}$, $2^{+}_{6}$, and $4^{+}_{6}$ states) might be the candidates
for the states observed recently in $^{10}$Be+$\alpha$ breakup reactions.
The $^{10}$Be+$\alpha$ correlation confirmed in the present study supports again this expectation.

We comment on the quantum fluctuation of the $\alpha$-$\alpha$ distances.
The full solution does not correspond to well defined distances
but to a superposition of the wave functions which have various distances
because the energy curves are very soft with respect to asymmetry and quantum fluctuation may be large.
However, because of the existence of minimum at the asymmetric point, components of asymmetric configuration 
are expected to largely mix in the superposed wave functions.
It means the importance of Be+$\alpha$ correlation.

Finally, we comment on the negative parity states of the linear-chain structures in $^{14}$C.
Because of the asymmetry of the intrinsic state from the $^{10}$Be+$\alpha$ correlation,
one can expect negative parity states as a parity partner of the positive-parity linear-chain states \cite{vonOertzen_14C_04}.
However, in the present calculations, energies of negative parity states are much higher, 
more than 10 MeV, than the positive-parity linear-chain states;
therefore, the negative-parity linear-chain states can not appear near the positive-parity linear-chain states.
It is consistent with the previous study on $^{14}$C with an AMD method \cite{Suhara_14C_10} which shows 
a negative-parity linear-chain state does not appear.

\section{Summary and outlook}\label{summary}

We investigated the linear-chain structures of $^{14}$C using a generalized molecular orbital model.
In this model, we incorporated asymmetric configurations of linear three $\alpha$ clusters 
to study the $^{10}$Be+$\alpha$ correlation in $^{14}$C.
We also investigated Be+$\alpha$ correlations in linear $3\alpha$ and $3\alpha$+$n$ systems 
to see the effect of three $\alpha$ clusters and that of valence neutron motion. 
The linear-chain structure of $^{16}$C is also studied. 

It was found that the linear-chain structure of $^{14}$C shows the $^{10}$Be+$\alpha$ correlation, which 
is characterized by the asymmetric $2\alpha$+$\alpha$ configuration 
with two valence neutrons distributing around the correlating $2\alpha$. 
Compared with the results for the linear $3\alpha$ and $3\alpha$+$n$ systems,
we find that a linear 3$\alpha$ system prefers the asymmetric 2$\alpha$+$\alpha$ configuration,
whose origin is the many body correlation incorporated by the parity projection.
This configuration causes an asymmetric mean field for two valence neutrons,
which induces the concentration of valence neutron wave functions around the correlating 2$\alpha$.
These are the origins of the $^{10}$Be+$\alpha$ correlation in the linear-chain structure of $^{14}$C.
In analysis of motion of valence neutrons,  
we find that both the kinetic energy gain and potential energy gain play important roles 
in the linear three $\alpha$ clusters with valence neutrons.
A symmetric configuration of linear three $\alpha$ clusters is indicated in $^{16}$C.
We also find valence neutrons in a linear-chain structure of $^{16}$C have multi-neutron correlation,
while those in $^{14}$C have no explicit two-neutron correlation.

As a future problem, it is important to clarify whether or not a linear-chain structure 
with a symmetric configuration appears in $^{16}$C.
For this aim, the AMD + GCM method is a promising framework for systematic study of 
$^{16}$C from low-lying states to high-lying states.

\section*{Acknowledgments}

The computational calculations of this work were performed  by supercomputers in YITP and KEK. 
This work was supported by the YIPQS program in YITP.
It was also supported by the Grant-in-Aid for the Global COE Program 
``The Next Generation of Physics, Spun from Universality and Emergence" 
from the Ministry of Education, Culture, Sports, Science and Technology (MEXT) of Japan,
and Grant-in-Aid for Scientific Research (Nos. 22$\cdot$84, 18540263, and 22540275) from JSPS.

\end{document}